\documentclass{PoS}

\usepackage{graphicx}
\usepackage{url}
\usepackage{amsmath}

\title{VERITAS monitoring of LS I +61$^\circ$ 303 in conjunction with X-ray, and GeV observation campaigns}

\ShortTitle{Multi-wavelength monitoring of LS I +61$^{\circ}$ 303}

\author{\speaker{Payel Kar} {for the VERITAS Collaboration}\thanks{https://veritas.sao.arizona.edu}\\
        University of Utah\\
        E-mail: \email{payel.kar@utah.edu}}

\abstract{One of the most enigmatic TeV binary systems, LS I +61$^\circ$ 303 exhibits a high degree of flux modulation from radio to TeV wavelengths over a single orbit of the binary system, once every $\sim$26.5 days. The binary system also exhibits a $\sim4.5$ year superorbital modulation in radio, X-ray and GeV emission which is yet to be seen in TeV gamma rays. LS I +61$^\circ$ 303 has been observed by VERITAS in the TeV energy range and multiwavelength partners (optical - GeV). The contemporaneous multiwavelength data sets enable searches for correlations between X-ray and TeV emission, as well as GeV and TeV emission. These correlation studies can further elucidate the astrophysical properties of this binary system. 
We present the preliminary results from the analysis of recent VERITAS observations from the 2014-2015 season in conjunction with \textit{Swift}-XRT (0.3-10 keV) and \textit{Fermi}-LAT (0.3-300 GeV) observations.}

\FullConference{The 34th International Cosmic Ray Conference,\\
		30 July- 6 August, 2015\\
		The Hague, The Netherlands}

\begin{document}

\section{Introduction}
The high-mass X-ray binary (HMXB) LS I +61$^{\circ}$ 303 is one of the most studied binary systems in the gamma-ray sky. It is located at a distance of $\sim2$ kpc \cite{frail} and consists of a massive B0 Ve star \cite{casares} and a compact object. The nature of the compact object, (either a black hole \cite{punsly,zimmerman} or a neutron star \cite{dhawan}) is a subject of active debate because of the poorly constrained mass of the compact object and uncertainties in the inclination angle of the system. No evidence for pulsed emission in the radio, X-ray, or GeV bands has been detected thus far. The orbital period  ($\sim26.5$ days \cite{gregory}) of the binary system modulates energetic outbursts in radio \cite{harrison}, X-ray \cite{greiner}, GeV \cite{abdo} and TeV wavelengths \cite{acciari,albert}. It has a highly elliptical orbit $(e = 0.54 \pm 0.03$ \cite{Aragona2009}) around the Be star which is losing mass due to its fast rotation, forming an equatorial disc \cite{casares}. The zero orbital phase $(\phi_o)$ is defined at MJD 43366, the date of the first radio detection of the system \cite{gregory}. Based on this the apastron passage occurs at an orbital phase $\phi=0.77$ and periastron passage at $\phi=0.27$ according to radial velocity studies \cite{Aragona2009}.

The system is also found to exhibit a superorbital period of $1667 \pm 8$ days in radio \cite{gregory}, $H_{\alpha}$ emission lines \cite{zamanov}, X-rays \cite{chernyakova} and GeV emission \cite{ackermann}. The source flares in X-ray wavelengths as seen by the RXTE satellite between $\phi\simeq0.35$ to $\phi\simeq0.75$ on the superorbital $\sim4.5$ years timescale. Radio wavelength bursts are also observed, delayed from the X-ray by $\Delta\phi \simeq 0.2$ \cite{chernyakova}. It is suggested that plasma bubbles consisting of energetic particles travel from a region close to the binary system to a radio emission region, located about 10 times the distance of separation between the members of the binary system. The time of flight for the plasma bubbles may cause delay of the radio emission \cite{chernyakova}. 

It was initially observed that the GeV flux as seen by \textit{Fermi}-LAT is also modulated by the orbital period with peak emission at $\phi \simeq 0.4$, near the periastron passage \cite{smith}. With further monitoring, it came to light that the periodic GeV emission at periastron occurs only during certain epochs. The lightcurve from \textit{Fermi}-LAT data shows a broader peak after periastron and a smaller peak just before apastron \cite{abdo}. Later timing analysis of the GeV lightcurve demonstrated the presence of periodicity but not with equal power at all times. There is a discontinuity in the periodicity of GeV emission at periastron and intermitent low flux just before apastron \cite{massi}. It is to be noted that this source exhibits different behavior in the low and high energy range. Although modulated by the superorbital period, the radio outbursts occurs consistently at every apastron passage.

LS I +61$^{\circ}$ 303 was first detected in the TeV energy range in 2006-2007 by MAGIC \cite{albert} and confirmed by VERITAS \cite{acciari,acciari2}. Initially, the source was detected in TeV over multiple orbits, with periodic outburts near apastron. Between 2008 and 2010 the source was not detected near the apastron passage but it was detected at a lower flux level during a single orbit near periastron passage \cite{aleksi}. Further monitoring in 2011-2012 season with VERITAS demonstrated LS I +61$^{\circ}$ 303 was active and flaring in the previously seen apastron region of the orbit \cite{smith}.

The aim of this paper is to present additional multiwavelength monitoring of the source in X-ray, GeV and TeV wavelengths by reporting the recent data taken in the 2014-2015 season.

\section{VERITAS Observations}
The \textbf{V}ery \textbf{E}nergetic \textbf{R}adiation \textbf{I}maging \textbf{T}elescope \textbf{A}rray \textbf{S}ystem consists of four 12m telescopes, each with a Davies-Cotton tessellated mirror structure, using the imaging atmospheric-Cherenkov technique to observe gamma rays from 85 GeV to > 30TeV. It is located at the Fred Lawrence Whipple Observatory (FLWO) in southern Arizona ($31^\circ 40'$ N, $110^\circ 57'$ W,  1.3 km a.s.l.). For detailed description and characterization of the instrument refer to \cite{park,staszak}.

\begin{figure}[!h]
    \centering
	\includegraphics[scale=0.45]{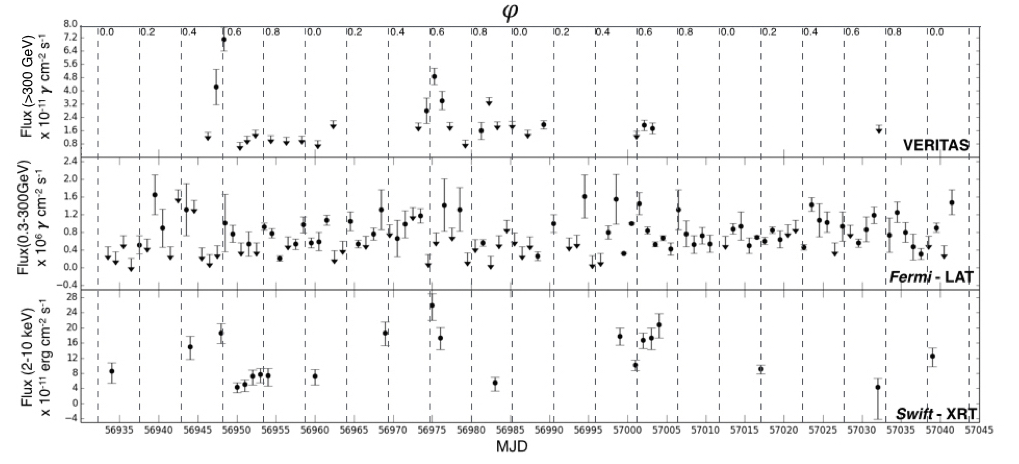}
	\caption{Daily light curves for LS I +61$^{\circ}$ 303 with VERITAS (>300 GeV), \textit{Fermi}-LAT (0.3-300 GeV) and \textit{Swift}-XRT (2-10 keV) during October 2014 - January 2015. The orbital phase $\phi$ of the system is also indicated. Flux upper limits for VERITAS are calculated at the $99\%$ confidence level for data points with significance < 3$\sigma$ and are marked as downward arrows. Flux upper limits for \textit{Fermi}-LAT are calculated at the $90\%$ confidence level and are marked similarly. }
	\label{lc}
\end{figure}

\begin{figure}
\centering
\begin{minipage}{0.4\textwidth}
\centering
	\includegraphics[scale=0.3]{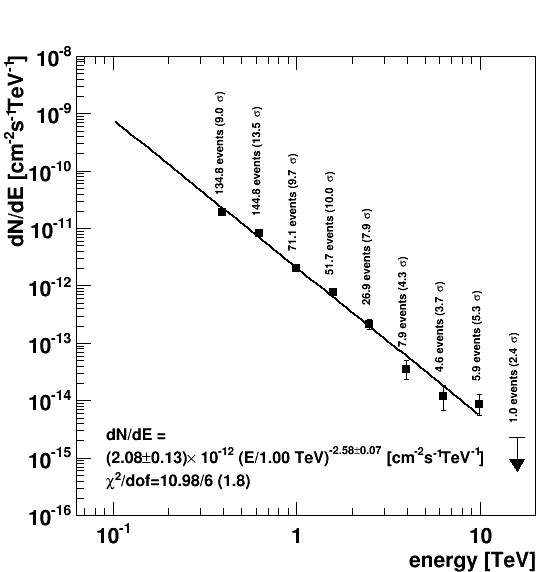}
	\caption{VERITAS Spectral Energy \mbox{Distribution $>300$GeV} during 2014-2015 observation season. The spectral index $(-2.58\pm0.07_{stat})$ is consistent with previous published value of $-2.59\pm0.15_{stat}$\cite{smith}. }
	\label{ed_spectra}
\end{minipage}\hfill
\begin{minipage}{0.5\textwidth}
	\medskip
	\includegraphics[scale=0.23]{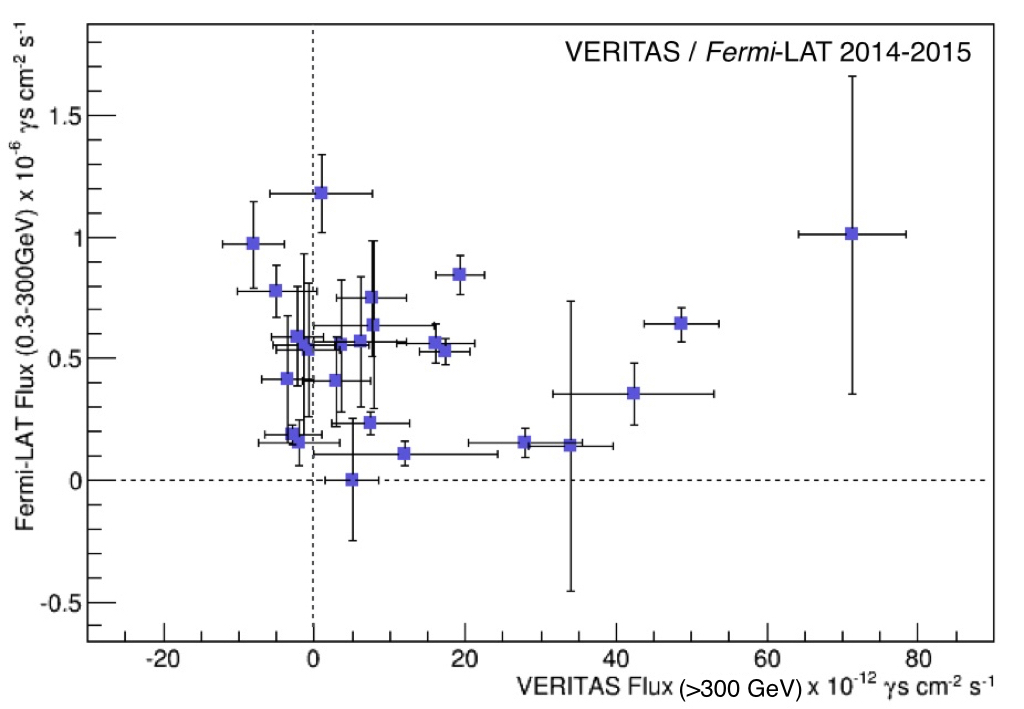}
	\medskip
	\caption{Daily flux correlation study for \mbox{VERITAS} and \textit{Fermi}-LAT. The Pearson correlation coefficient of $0.071^{+0.36}_{-0.39}$ indicates independent data sets. It is consistent with previously published correlation \mbox{studies}. }
	\label{gev_tev_corr}
\end{minipage}
\end{figure}

VERITAS has been monitoring LS I +61$^\circ$ 303 every year since 2007. The data presented here were recorded from October 16 2014 to January 10 2015 (MJD 56946 to MJD 57032) amounting to a total of $\sim28$ hours of quality-selected livetime observations. The data is unevenly sampled around the apastron passage between $\phi\sim0.4$ to $\phi\sim0.9$. During the first two orbital phases observed in October and November the source was in an extremely active state producing a flux of $15-70 \times 10^{-12}\: \gamma$s cm$^{-2}$ s$^{-1}$ above 300 GeV corresponding to 25$\%$ of the Crab Nebula flux in the same energy range. The lightcurve of this source can be seen in Figure \ref{lc}. This was historically the highest flux recorded ever for this binary system in VERITAS. An excess of 563 events over the background was detected corresponding to a statistical significance of $\sim19\sigma$. The source was found to revert to its typical flux level $(\sim 15\%$ of the Crab Nebula flux) during the last observed orbit in December 2014. As shown in Figure \ref{ed_spectra}, the  differential energy spectrum in the range 0.15-10 TeV is well fitted to a power-law model $(\chi^{2}/n.d.f.=10.98/6)$ represented by $(2.08\pm0.13_{stat}) \times 10^{-12} \times (\frac{E}{1.00 TeV})^{(-2.58\pm0.07_{stat})} \gamma s \:cm^{-2} s^{-1} TeV^{-1}$. The energy spectrum is found to be consistent with previous observations at a lower flux level \cite{acciari,smith,aleksi}. Nightly variabilty is also observed at the level of $3\sigma$ level as described in \cite{anna} following the method in \cite{smith}.

\section{\textit{Fermi}-LAT Observations}
The \textit{Fermi} Gamma-ray Space Telescope \cite{atwood} has two components, the Gamma-ray Burst Monitor (GBM) dedicated to the observation of GRBs and the Large Area Telescope (LAT) which has a large field of view for gamma rays in the energy range of 100 MeV - 500 GeV. The analysis was performed on all data available between October 3rd 2014 (MJD 56933) and January 18th 2015 (MJD 57040) using the Fermi Science tools  v10r0p5 available from the Fermi Science Support Center (FSSC).\footnotemark[1] \footnotetext[1]{\url{http://fermi.gsfc.nasa.gov/ssc/data/analysis/software/}} Only "SOURCE" class events (event class 128) were analyzed from Pass 8 data with event reconstruction following the standard quality cuts prescribed by FSSC. Other standard cuts applied are for zenith angle < $90^{\circ}$ to reduce gamma rays from the Earth's limb. A region of interest (ROI) of $15^{\circ}$ was initially chosen and all 3FGL sources within it were taken into account in the fit. The spectral parameters of all sources beyond $5^{\circ}$ from LS I +61$^\circ$ 303 were fixed and any source with a TS < 2 was removed from the model. This model was used to produce daily lightcurves by fixing all sources except LS I 61$^{\circ}$ + 303 and the nearby pulsar 3FGL J0248.1+6021.

Spectral analysis was performed using likeSED (a user contributed tool) from FSCC.\footnotemark[2]  The spectrum of the source is fitted with a power law with exponential cutoff of the form \begin{equation} N_0(\frac{E}{1 MeV})^{-\Gamma} e^{(-E/E_{cutoff})} \end{equation} where N$_0=(9.5 \pm 3.1) \times 10^{-5} \: \gamma s$ MeV$^{-1}$ cm$^{-2}$ s$^{-1}$, \mbox{ $\Gamma=1.97 \pm 0.05$,} and E$_{cutoff}=4.00\pm0.69$ GeV.

The correlation study for GeV and TeV flux shown in Figure \ref{gev_tev_corr}, yielded a Pearson correlation coefficient of $0.071^{+0.36}_{-0.39}$ indicating independent data sets as previously observed \cite{smith}.

\section{\textit{Swift}-XRT Observations}
\textit{Swift} \cite{burrows} is a spaced-based telescope primarily involved in observing GRBs. It carries three payloads of which the X-ray Telescope (XRT) is capable of imaging and obtaining spectra for pointed observations in the energy range of 0.3-10 keV. It has a flexible Target of Opportunity program which supports multiwavelength observations. VERITAS made several simultaneous observations with \textit{Swift}-XRT for LS I +61$^\circ$ 303 during the 2014-2015 season. The \textit{Swift}-XRT data has been analysed using the webtools available from UK Swift Science Data Centre\footnotemark[3] \footnotetext[3]{\url{http://www.swift.ac.uk/user_objects/}}. The data is processed to create X-ray light curves and spectra based on HEAsoft v6.16 package.

\begin{figure}
    \centering
	\includegraphics[scale=0.2]{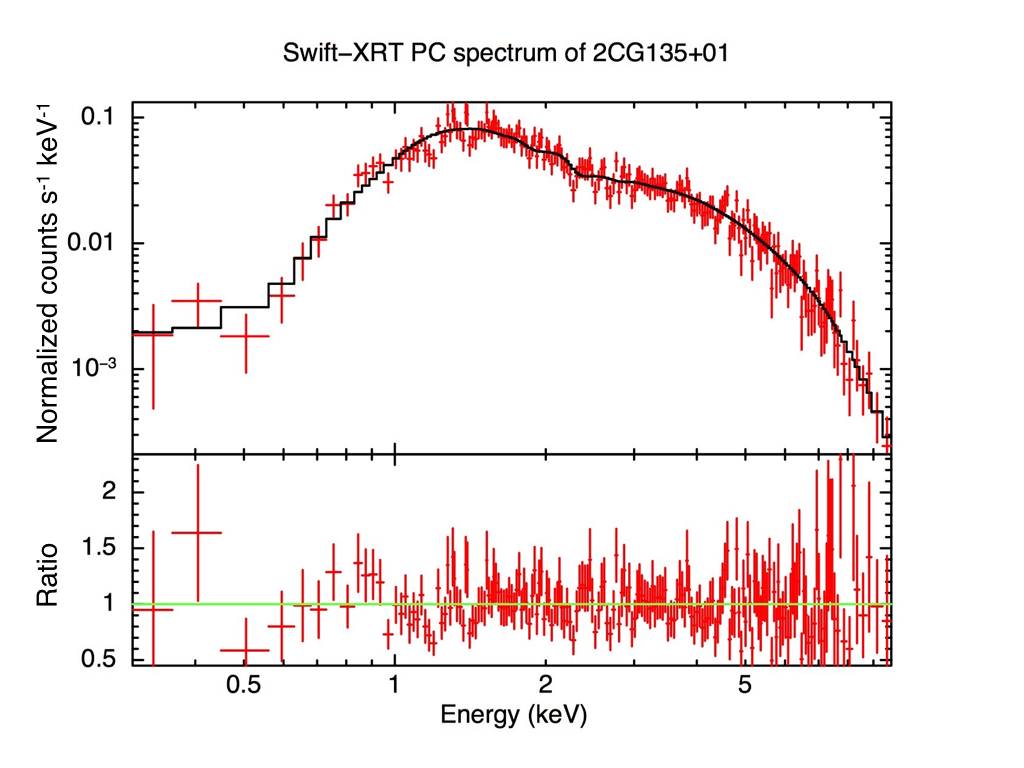}
	\caption{\textit{Swift} X-ray spectrum of LS I +61$^\circ$ 303 fits a power law model with photoelectric absorption.}
	\label{xrt_spectra}
\end{figure}

\begin{figure}[!ht]
    \centering
	\includegraphics[scale=0.4]{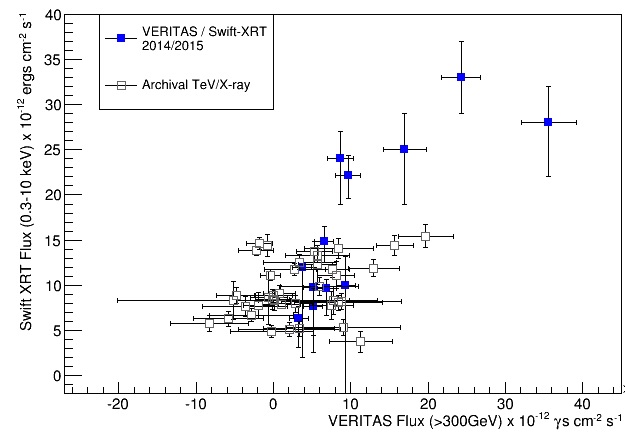}
	\caption{Correlation study of contemporaneous \textit{Swift}-XRT and VERITAS flux data points. A Pearson correlation coefficient of $0.80^{+0.14}_{-0.38}$ indicates a correlation between the X-ray and VHE gamma-ray fluxes. Archival data, shown as open squares showed no evidence of correlation with a correlation coefficient of $0.36\pm0.32$ \cite{smith}. }
	\label{xray_tev}
\end{figure}

The XRT data was recorded between October 4 2014 and January 17 2015 (MJD 56934 - MJD 57039). The source spectrum for the complete data set is shown in Figure \ref{xrt_spectra}. The best fit simple power-law model with photoelectric absorption gives a photon index of $1.54^{+0.08}_{-0.07}$ with $\chi^2/n.d.f.= 611.59/700$ and neutral hydrogen density of $0.87^{+0.08}_{-0.08} \times 10^{22}\:$ cm$^{-2}$. The photon index was found to vary from $1.3^{+0.7}_{-0.6}$ to $1.8^{+0.6}_{-0.5}$ with a reduced-$\chi^2$ value ranging from 0.71 to 1.15 for individual observations. VERITAS recorded 12 contemporaneous observations with \textit{Swift}-XRT which are marked on the Figure \ref{xray_tev}. Archival data \cite{smith} from 2011/2012 and previous VERITAS and MAGIC measurements \cite{acciari3} are also shown in this figure. A Pearson correlation coefficient of $0.80^{+0.14}_{-0.38}$ between \textit{Swift}-XRT and VERITAS fluxes was found for the current 2014-2015 data. The total correlation coefficient considering the current data and archival data was found to be $0.67^{+0.18}_{0.14}$.

\section{Results and Discussion}
The results for the multiwavelength monitoring of LS I +61$^\circ$ 303 have been presented for 2014-2015. The system was exceptionally bright in TeV during the first two observed orbital cycles in October and November 2014 and then retreated to its quiescent flux levels during December 2014. A correlation study performed with GeV and TeV gamma-ray fluxes confirmed previous results indicating a lack of correlation. On the other hand, an indication of a correlation is seen between the X-ray and TeV data sets. No significant variation of the spectral indices has been found in X-ray, GeV or TeV gamma rays. 

\begin{figure}[!h]
    \centering
	\includegraphics[scale=0.3]{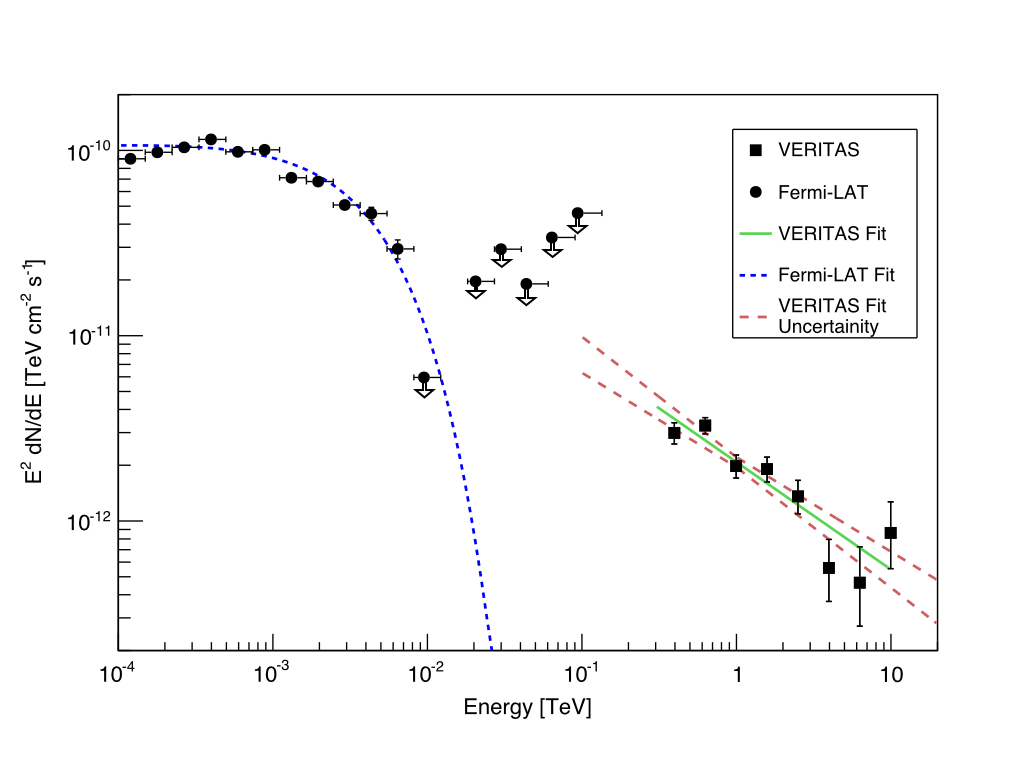}
	\caption{VERITAS and \textit{Fermi}-LAT Spectral Energy Distribution}
	\label{gev_tev_sed}
\end{figure}

A detailed GeV-TeV spectral energy distribution (SED), seen in Fig \ref{gev_tev_sed} is obtained from \textit{Fermi}-LAT and VERITAS data for 2014-2015 season. Similar to published observations \cite{smith}, this contemporaneous SED demonstrates a cutoff at $\sim 4$ GeV. No emission is detected between 10-300 GeV. Below 10 GeV, the data is fitted with a power-law with exponential cutoff. A simple power-law is fitted to the data above 300 GeV. The gap in detected emission may be indicative of GeV and TeV gamma rays originating from different particle populations. This hypothesis of different parent particle population is also supported by the lack of correlation in GeV and TeV fluxes shown in Fig \ref{gev_tev_corr}.

A potential correlation seen between the X-ray and TeV gamma ray data suggest that emission in these two energies may be produced by the same population of accelerated particles. Leptonic models may be a viable option in such a case as suggested previously \cite{aleksi}. Synchrotron radiation from high energy electrons may drive the X-ray emission and these electrons may later undergo Inverse Compton scattering, upscattering the photons of the companion star to very high energies. A similar stronger correlation was seen during a flaring episode in the September of 2007 between VHE data from MAGIC and X-ray data from XMM-Newton and Swift \cite{anderhub}. It was postulated that uncertainties make it difficult to draw conclusions about correlations outside outbursts unless simultaneous observations were recorded. The historically brightest flare of LS I +61$^\circ$ 303 seen in the 2014-2015 season by VERITAS bolsters the previously found correlation between X-rays and TeV gamma rays during September 2007 flaring episode.

\medskip
\medskip
\noindent\textbf{Acknowledgments}

This work made use of data supplied by the UK Swift Science Data Centre at the University of Leicester.
This research is supported by grants from the U.S. Department of Energy Office of Science, the U.S. National Science Foundation and the Smithsonian Institution, and by NSERC in Canada. We acknowledge the excellent work of the technical support staff at the Fred Lawrence Whipple Observatory and at the collaborating institutions in the construction and operation of the instrument. The VERITAS Collaboration is grateful to Trevor Weekes for his seminal contributions and leadership in the field of VHE gamma-ray astrophysics, which made this study possible.

\bibliography{skeleton}

\end{document}